# OLAP on Structurally Significant Data in Graphs

Kifayat Ullah Khan[1], Kamran Najeebullah[2], Waqas Nawaz[3], Young-Koo Lee[4]
Department of Computer Engineering, Kyung Hee University
Yongin-si, Gyeonggi-do 446-701, Republic of Korea
kualizai@hotmail.com[1], applelogix.kamran@gmail.com[2], {wicky786[3],yklee[4]}@khu.ac.kr

## ABSTRACT
Summarized data analysis of graphs using OLAP (Online Analytical Processing) is very popular these days. However due to high dimensionality and large size, it is not easy to decide which data should be aggregated for OLAP analysis. Though iceberg cubing is useful, but it is unaware of the significance of dimensional values with respect to the structure of the graph. In this paper, we propose a Structural Significance, SS, measure to identify the structurally significant dimensional values in each dimension. This leads to structure aware pruning. We then propose an algorithm, iGraphCubing, to compute the graph cube to analyze the structurally significant data using the proposed measure. We evaluated the proposed ideas on real and synthetic data sets and observed very encouraging results.

## 1. INTRODUCTION
Graphs, like social networks, are growing rapidly in size. As a result of this, their summarized analysis is very popular these days to view the data from multiple dimensions and at various granularity levels. In graphs, the dimensions are the attributes attached to nodes. Graph Cube [1] is a very useful option in this scenario. It lays the foundation to perform OLAP on graphs for analyzing the structural relationships among various dimensional values (DVs i.e. aggregate nodes in each aggregate graph) in each dimension.

Though the Graph Cube facilitates OLAP on graphs, it is not necessary that the analyst is interested in all aggregations of the underlying graph. For example, if she is only interested in analyzing the relationships among the high profile authors in the DBLP co-authorship network, then computing the relationships of low profile authors is useless. One opportunity is to create an iceberg cube to aggregate only those DVs in each dimension which satisfy minimum support threshold. On the other hand, using such criteria in case of graphs, prunes the DVs without considering their structural significance. For instance the DV C2 attached to nodes 2 and 4 in Fig. 1, has high degree, betweenness and closeness centralities but it is pruned if the support count is say 3. The motivation is that graph is about relationships analysis so the criteria must consider all the features affecting the relationships in it. In this regard, we propose a Structural Significance, SS, measure which finds the structural significance of each DV in each dimension.

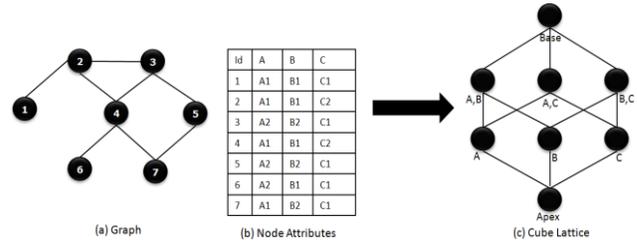

**Figure 1. Illustrating Cube Computation Process for 3-Dimensional Graph**

When the structurally significant DVs in the graph are identified, we need to compute the cube to analyze the summarized relationships. As there exists a plethora of relational data cube computation algorithms, any algorithm can be utilized with minor modifications [1]. However a common problem in most of them is that they suffer from the curse of dimensionality. In [2] authors used inverted index based small data fragments and proposed to compute the local data cubes for each fragment to solve this problem. Their proposed algorithm follows the computation order of [3] which is sensitive to high cardinality and order of dimensions. Considering these issues, we propose an algorithm, iGraphCubing, to overcome the problems of existing algorithms.

Following are the contributions of this paper

(a) Structure Aware Pruning criteria to distinguish the structurally significant DVs in each dimension of multidimensional networks. It focuses the structural significance of each DV which is over shadowed by high minimum support.

(b) A cube computation algorithm, iGraphCubing, for graphs (multidimensional networks.) iGraphCubing uses inverted indices as underlying data format to compute the graph cube in bottom-up and breadth first style and computes the multiple granularity levels of cube lattice using 2n-Steps-Up Aggregations.

(c) Evaluation of the proposed ideas on various real world datasets.

Rest of the paper is organized as follows. Section 2 explains the proposed structure aware pruning measure, also detailing the existing graph structural measures and their limitations. Section 3 presents the proposed cube computation algorithm which is followed by related work in section 4. Section 5 lists the experiments which are followed by conclusion in section 6.

## 2. STRUCTURE AWARE PRUNING

In this section, we first describe various categories of existing graph structural measures along with their limitations and then we present our proposed measure to perform structure aware pruning.

### 2.1 Existing Graph Measures as Criteria to Determine Significance

There exist a number of graph structural measures for various application purposes [4]. They can be classified into Degree Measures, Distance Measures, Connectivity Measures, Reciprocity and Transitivity measures, Centrality Measures and Homophily, Assortative Mixing, Similarity measures. Each category further contains a list of measures. These categories can be classified into higher level categories, as displayed in Table 1. From this listing, the structure based measures are easily located to be used to find the significance of each dimensional value.

**Table 1. Categorization of Graph Structural Measures**

| Category | Constituting Measures |
|---|---|
| Distance | Distance measures |
| Similarity | Homophily, Assortative Mixing, Similarity measures |
| Location | Connectivity Measures, Centrality Measures |
| Structure | Degree Measures, Reciprocity and Transitivity |

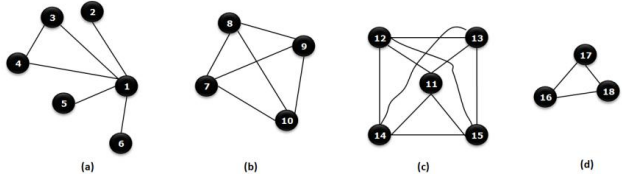

Figure2. Sample Graphs Describing Various Structural Measures

### 2.2 Limitations of Existing Measures

From the categorization of graph measures in Table 1, structure oriented measures are apparent. Now we evaluate these measures to find the significance of DVs. We consider the degree and clustering co-efficient as the representative from each type. The values of these two measures are assigned to each DV of all the vertices to find their significance.

Vertex 1 in Figure 2 (a) has degree higher than vertex 7 in (b) but it has lower density than that of vertex 7. So we cannot use degree as the significance criteria. Similarly vertex 11 in (c) and 16 in (d) has same clustering co-efficient though neighboring sub-graph around vertex 11 is much denser than that of 16. Hence it depicts the limitation of clustering co-efficient. Nevertheless all these measures have their own importance however they are unable to stand true when the requirements are slightly changed.

As graph cube is about analyzing the relationships among DVs, so the criteria to identify the significance of each DV must be comprehensive in terms of having impact on the structure of the underlying graph.

### 2.3 Structural Significance

The proposed Structural Significance, SS, measure, identifies the significance of each DV by inspecting its impact on the structure of the graph. There are three main components of the proposed measure. (a) The diversity of the neighboring attribute values of each vertex, (b) clustering co-efficient and (c) the density around each vertex in the graph. We believe that the above three factors are affected due to DVs of each vertex i.e. certain DVs have great impact on above factors while other have medium to least. We assign all three measure values of each vertex to all the DVs of vertex under consideration. In this way, the significance of each DV is calculated. Equation 1 presents the significance calculation of each DV.

$$SS(a_{j_i}) = \sum_{j=1}^{n}\sum_{i=1}^{m}(\alpha \cdot CC(a_{j_i}) + Density(a_{j_i})) \quad (1)$$

Where $SS(a_{j_i})$ denotes the significance of jth attribute, $1 \leq j \leq m$, of vertex i, $1 \leq i \leq n$, n is the total number of vertices in the graph. α is neighborhood attributes diversity, telling how diverse the attribute values of the adjacent nodes are, and CC is clustering co-efficient. Using the above equation, the significance of each DV with respect to every node in the graph is calculated. All the DVs below average significance are pruned, which depicts structure aware pruning.

## 3. CUBE COMPUTATION

In this section, we explain the proposed algorithm. It operates on the graph stored in inverted index format. This format facilitates bottom-up cube computation to prune irrelevant data due to anti-monotonicity property, to compute child aggregations from their parents rather than relying on the source data and curse of dimensionality. Moreover it moves up the cube lattice in level by level manner, where there exists the room to accelerate the computation process.

We first present the Level-by-Level process, followed by an acceleration during the course of action. We then present the algorithm, iGraphCubing, to compute the cube. Finally we explain how the aggregate nodes having invalid labels are avoided.

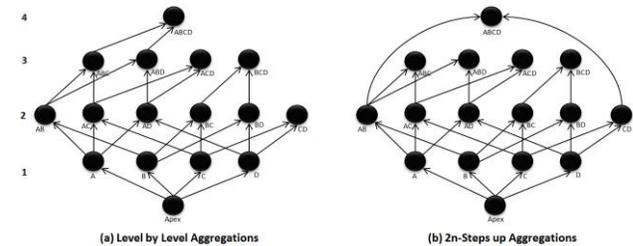

Figure 3. iGraphCubing Computation Cycle

### 3.1 Level by Level Aggregations

Figure 3 displays the Level-by-Level computation cycle. The first level of aggregate networks, e.g. A, B, C, D highlighted with the label 1, is computed from the underlying graph and is stored. The second level of aggregations is computed from already computed first level aggregations. In this all the aggregate networks at level n are computed from their parents at level n-1.

### 3.2 2n-Steps Up Aggregations

Computing the SiG Cube in Level-by-Level manner has the room to accelerate the computation process. For example, during the computations of the aggregate networks above the second level of aggregations, i.e. $AG_{ijk}$ from $AG_{ij}$ and $AG_{jk}$, there exists a common dimension e.g. dimensions j. This common dimension happens to be a bridge to compute the next level of aggregations. However the situation in which there is no common dimension(s), i.e. $AG_{ij}$ and $AG_{kl}$, the node and edge ids contribute to compute the aggregation $AG_{ijkl}$, which is a step-up in the granularity level of cube lattice. Such step-up avoids computations at the level, to compute its successor/child level, for which all the aggregations have been pre-computed. For example, when the algorithm is computing the $3^{rd}$ level aggregate networks from $2^{nd}$ level, the aggregate networks at $4^{th}$ level are computed at the same time. So there is no need to explicitly move to the $3^{rd}$ level to compute the $4^{th}$ level aggregate networks. Intuitively we can observe that at the level n, all the aggregations till the level 2n can be computed. We term it as 2n-Steps Up aggregation which results in significant acceleration in cube computation process. Figure 7 (b) displays the 2n-Steps Up computation process.

## 3.3 Handling Aggregate Nodes with Invalid Labels

As explained, to compute the nodes for an aggregation level n, the nodes at n-1 level are used. In this process, there is the possibility of creating nodes having invalid labels e.g. ABC is valid but ACB is not. To avoid such situation, we utilize the concept of Light Weight Signatures (LWS). The signatures of any cuboid in the cube lattice are light weight, e.g. ABC, while the valid signatures of any cell in the cube are termed as heavy weight signatures, e.g. $a_1b_2c_1$. The heavy weight signatures are not feasible as they are proportional to the cardinality of each dimension.

LWS can be generated for the cube lattice using a data generator, taking all the dimensions as input. However this involves computation and comparison overhead plus memory requirement to check for each newly created cell. One opportunity is to rename the dimension names into chronological list of alphabets and compare the ASCII code for each alphabet in the label of newly created node.

## 3.4 iGraphCubing, the proposed algorithm

The graph cube consists of aggregate networks at granularity levels equal to the number of vertex centric dimensions in the underlying graph. Each aggregate network is either 1-dimensional or n-dimensional containing corresponding aggregate nodes, self edges and cross edges. So the main theme of iGraphCubing is node aggregation and edge aggregation in the aggregate networks. The process of edge aggregation is similar to that of node aggregation so we skip it for the sake of brevity and explain only the node aggregation process.

Algorithm 1 outlines the node aggregation process. It receives as input the graph G, SAP to prune insignificant dimensional values and the level Lv for which to compute the aggregate network.

Line 2 iterates through the DVs of all the dimensions of the underlying graph G for aggregate networks of level 1. A new node is created, for level 1 aggregate network at line 3, with label of DV and having a list of node ids which share the same DV in G. The significance of each dimensional value becomes the significance of this newly created node and is matched against SAP at line 4.

**Algorithm 1** Node Aggregation

Input: Graph G, Significance SS, Level $Lv$
Globals: Light Weight Signatures $flagLWS$, NodeList NL, Dimensional Value DV
Output: Aggregate Nodes of each granularity level

1: **if** ($Lv = 1$)
2:    **for each** DV in $\{D_1, \ldots, D_n\}$ of G
3:       $N_i \leftarrow DV\ Label, NodeList(DV)$
4:       **if** (Significance($N_i$) $\geq SAP$)
5:          output $N_i$
6: **else**
7:    **for each** $N_i, N_j$ of Previous Level Data
8:       **if** ($Dimension(N_i) \neq Dimension(N_j)$)
9:          $NL_i \leftarrow NodeList(N_i)$
10:         $NL_j \leftarrow NodeList(N_j)$
11:         $N_k \leftarrow NL_i \cap NL_j$, Label ($N_i \cup N_j$)
12:         **if** (Lv ($N_k$) $\geq 2$ && Lv ($N_k$) $< 2n$ && $flagLWS = true$)
13:            output $N_k$
14:         **else if** (Lv ($N_k$) $= 2n$ && $flagLWS = true$)
15:            output $N_k$ as extra level node
16: Mark respective nodes in iGraph as Previous Level Nodes

If it satisfies the criteria, it is output at line 5. To compute the 2-dimensional nodes, the previous level data is 1-dimensional list of nodes. Line 7 iterates through all these nodes, considering two nodes at a time. If the dimensions of both of them do not match (line 8) then the node id lists of each of them are retrieved at lines 9 and 10. The intersection of node ids and union of their labels is performed at line 11 to create the new node. If the nodes are being computed for the second level of aggregations and similarly the computed label is verified using light weight signatures, at line 12, then the node is output at line 13. In this way all the 2-dimensional nodes are computed. If the total number of dimension is, say 4, the computes nodes are marked as previous level data at line 16 as they will be used to compute the higher level of aggregations. In this case, when $3^{rd}$ level of nodes is being computed, the condition at line 12 is satisfied. Here 2n is 4 as the level being used, to compute the next level nodes, is 2. However as soon as the union of node labels appears to be of length 4 at line 11, the condition at line 14 is satisfied provided the intersection does result in some node ids and light weight signatures are correct. These nodes are output and are marked as extra level nodes. Such nodes are the result of the 2n-Steps up aggregations. As there are totally 4 dimensions, so the process stops otherwise these nodes serve to compute its successors.

## 4. RELATED WORK

In this section we present the related work with respect to the contributions of our paper i.e. (a) OLAP on Graphs, (b) Pruning and (c) cube computation.

OLAP on graphs and multidimensional networks is becoming very popular these days. It provides the opportunity to analyze the aggregate data from multiple dimensions and at various level of granularity. Graph OLAP [5] and Graph Cube [1] models are the pioneers to introduce OLAP on graphs They lay down the mechanism to define dimensions and measures in case of graphs. In this research work we target OLAP on single large multidimensional network. We focus the structural significance of the aggregate vertices which is not the focus the previous approaches. Identifying and computing the cube for structurally significant dimensional values in each dimension, help perform analysis of data which has impact on the structure of the data.

[6] Provides interestingness measure to identify the set of most useful summaries of data out of a large number of summaries. On the other hand, our proposed criteria focus the significance of each dimensional value having significant role in the structure of the network.

With regards to the cube computation, a number of algorithms exist. Top-down [7], Bottom-up [3] and integrated method [8] are the most prominent ones. Though the above methods are efficient in their own problem domains, however they do not focus the high dimensionality problem. The authors in [2] proposed an inverted indexed based method to overcome this issue. Our proposed method is based on this idea in order to efficiently compute the cube of the graph data.

## 5. EXPERIMENTS
In this section, we briefly provide the effectiveness and efficiency evaluations along with the datasets used.

### 5.1 Effectiveness Evaluation
We used real world dataset of the most popular online social network, Pokec, in Slovakia [9]. The retrieved nodes and edges are 2,37,604 and 14,52,811 respectively where each node is attached with 9 attributes. Figure 4 displays the significance and support count of each attribute value and depicts that in most cases support count donot have any impact on the significance of attribute values and vice verca, as highlited.

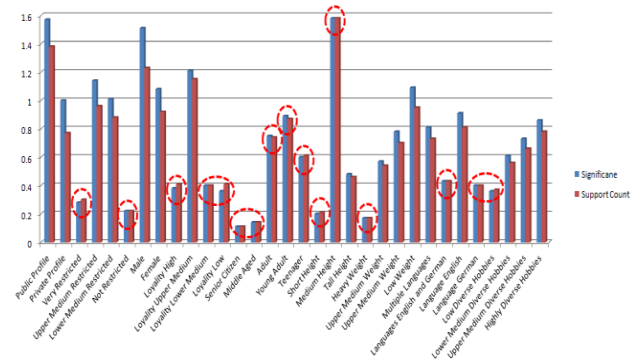

**Figure4. Significance and Support Count for attributes in Pokec Data**

### 5.2 Efficiency Evaluation
We used DBLP data set [10] for efficiency evaluation. The nodes in this version of DBLP dataset are not attached with any attributes, so we randomly generated 6 attributes to analyze the efficiency of the proposed algorithm. Figure 4 displays the efficient execution of 2n-Steps Up Aggregations compared to Level-by-Level aggregations.

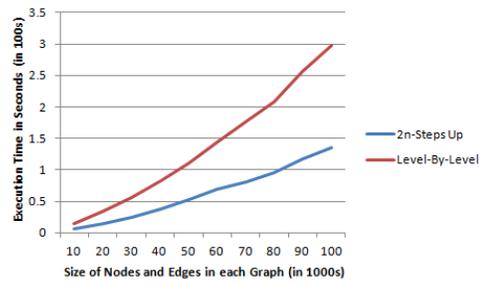

**Figure 5. Execution time on DBLP Data**

## 6. CONCLUSION
Considering the popularity of OLAP on graphs, we propose a Structure Aware Pruning measure and an algorithm to compute the cube for graphs.

As a future work, we plan to (a) broaden the scope of proposed measure till n hops and to (b) perform sub-graph structure based aggregations in the proposed algorithm.

## 7. ACKNOWLEDGMENTS
Our thanks to ACM SIGCHI for allowing us to modify templates they had developed.